\documentclass[12pt]{article}
\begin{document}
\footnotesize{PARTIAL DIFFERENTIAL EQUATIONS/EQUATIONS AUX DERIVEES PARTIELLES\\
MATHEMATICAL PHYS/PHYSIQUE MATHEMATIQUE
\begin{center}\bf
\Large{Probl\`emes de Cauchy avec des conditions modifi\'ees pour les \'equations d'Euler-Poisson- Darboux}
\end{center}
\begin{center}
Cheikh Ould Mohamed El-hafedh et Mohamed Vall Ould Moustapha
\end{center}
\begin{center}\bf
\large{Cauchy problems with the modified conditions for the Euler-Poisson-Darboux equations}
\end{center}
\begin{flushleft}
\scriptsize{{\bf Abstract}.	
Nowadays the Euler-Poisson-Darboux equation is extensively studied in several settings. The main questions on every spaces are explicit solutions for the classical Cauchy problems with the second data null. In this note we will generalize and unify several results on Euler-Poisson-Darboux equation. We consider the Cauchy problems with modified conditions for the classical and radial Euler-Poisson-Darboux equations. We give the explicit solutions in terms of the Gauss $_{2}F_{1}$ and Appell $F_{4}$ hypergeometric functions. The main results have many applications such as the classical and radial wave equation as well as the Tricomi operator [1].} 
\end{flushleft}
{\bf 1. INTRODUCTION}. On consid\`ere la famille classique d'\'equations d'Euler-Poisson-Darboux dans $R^{n}$
$$\Delta_{n} U(t,x)=(\frac{\partial^{2}}{\partial t^{2}}+ \frac{1-2\mu}{t}\frac{\partial}{\partial t})U(t,x),~~~~~t>0~~~~~~~(E_{1})$$
et la famille radiale d'\'equations d'Euler-Poisson-Darboux
$$(\frac{\partial^{2}}{\partial x^{2}}+\frac{1-2\nu}{x}\frac{\partial}{\partial x})U(t,x)=(\frac{\partial^{2}}{\partial t^{2}}+ \frac{1-2\mu}{t}\frac{\partial}{\partial t})U(t,x),~~~~~t>0,~x>0~~~~(E_{2})$$
avec les conditions initiales modifi\'ees
$$U\left(0,x\right)=f\left(x\right),~~~~\lim_{t\rightarrow0}t^{1-2\mu}\frac{\partial}{\partial t}U\left(t,x\right)=g\left(x\right)~(C_{1})$$
$$U\left(0,x\right)=A^{q}_{x}f\left(x\right),~\lim_{t\rightarrow0}t^{1-2\mu}\frac{\partial}{\partial t}U\left(t,x\right)=A^{q}_{x}g\left(x\right)~(C_{2})$$
o\`u $A^{q}_{x}$ est la $q^{\mbox{i\`eme}}$ puissance de l'op\'erateur
$A_{x}=\left\{
\begin{array}{rr}
\Delta_{n}~\mbox{si}~x\in R^{n}\\
\Lambda_{x}~\mbox{si}~x\in R^{+}\\
\end{array}\right.$,\\
$\Delta_{n}=\frac{\partial^{2}}{\partial x^{2}_{1}}+\frac{\partial^{2}}{\partial x^{2}_{2}}+...+\frac{\partial^{2}}{\partial x^{2}_{n}}$ et $\Lambda_{x}=\frac{\partial^{2}}{\partial x^{2}}+\frac{1-2\nu}{x}\frac{\partial}{\partial x},~\nu,~\mu$ et $q$ sont des param\`etres r\'eels.\\ L'\'equation $(E_{1})$ a \'et\'e \'etudi\'ee pour les valeurs enti\`eres de $k,~(1-2\mu=k)$ par A.Weinstein [9] et son \'ecole de Maryland, D.W.Bresters $[3]$ a exprim\'e la solution de l'\'equation $(E_{1})$ avec les conditions
$$U(0,x)=f(x),~U_{t}(0,x)=g(x)~~~~~~(C)$$
Mais il a \'et\'e contraint -semble t-il- de prendre la deuxi\`eme donn\'ee nulle $(g=0)$ \`a cause de la singularit\'e en $t=0$, les conditions modifi\'ees $(C_{1})$ et $(C_{2})$ permettent de rem\'edier \`a ce probl\`eme et de pouvoir prendre la deuxi\`eme donn\'ee  une fonction non nulle $g$, tout en recouvrant les conditions classiques $(C)$: ainsi pour $\mu=\frac{1}{2}$ on retrouve les probl\`emes de Cauchy pour les \'equations classiques et radiales des ondes (voir $[4]$ et $[2]$).\\ 
$(E_{1})$ et $(E_{2})$ sont des \'equations des ondes avec potentiels d\'ependants du temps  respectivement: $-\frac{1-2\mu}{t}\frac{\partial}{\partial t}$ et $\frac{1-2\nu}{x}\frac{\partial}{\partial x}-\frac{1-2\mu}{t}\frac{\partial}{\partial t}$.\\ 
L'inter\^et des \'equations $(E_{1})$ et $(E_{2})$ vient du fait que les potentiels correspondants sont homog\`enes de degr\'e $-2$ et donc les op\'erateurs de gauche et droite se comportent de la m\^eme mani\`ere. L'une des difficult\'es majeures dans le cas du potentiel d\'ependant du temps est l'absence de relation entre les semi-groupes engendr\'es par l' \'equation de Schrodinger et les propri\'et\'es spectrales de l'op\'erateur $H=-\Delta+V$. Rappelons que pour un potentiel ind\'ependant du temps $V$ on a $g(H)=\int g(\lambda)dE(\lambda)f$ o\`u $dE(\lambda)$ est la mesure spectrale associ\'ee \`a l'op\'erateur H; ce qui n'est pas valable dans le cas d'un potentiel d\'ependant du temps.
Les r\'esultats principeaux de cet article sont les suivants:\\
\\
{\bf TH\'EOR\`EME 1}. Pour $0<\mu<\frac{1}{2}$, le probl\`eme de Cauchy $(E_{1}),~(C_{1})$ admet la solution unique donn\'ee par:\\
\\
$U\left(t,x\right)=\alpha_{n,-\mu}t^{2\mu}\left(\frac{\partial}{t\partial t}\right)^{\frac{n-1}{2}}\int_{\left|x'-x\right|< t}f\left(x'\right)\left(t^{2}-\left|x'-x\right|^{2}\right)^{-\mu-\frac{1}{2}}dx'\\+\frac{1}{2\mu}\alpha_{n,\mu}\left(\frac{\partial}{t\partial t}\right)^{\frac{n-1}{2}}\int_{\left|x'-x\right|< t}g\left(x'\right)\left(t^{2}-\left|x'-x\right|^{2}\right)^{\mu-\frac{1}{2}}dx'$ si $n$ est impair,\\
\\
$U\left(t,x\right)=\beta_{n}t^{2\mu}\left(\frac{\partial}{t\partial t}\right)^{\frac{n}{2}}\int_{\left|x'-x\right|< t}f\left(x'\right)\left(t^{2}-\left|x'-x\right|^{2}\right)^{-\mu}dx'\\+\frac{1}{2\mu}\beta_{n}\left(\frac{\partial}{t\partial t}\right)^{\frac{n}{2}}\int_{\left|x'-x\right|< t}g\left(x'\right)\left(t^{2}-\left|x'-x\right|^{2}\right)^{\mu}dx'$
si $n$ est pair\\
avec $\alpha_{n,\mu}=\frac{\Gamma\left(1+\mu\right)}{2^{\frac{n-1}{2}}\pi^{\frac{n}{2}}\Gamma\left(\frac{1}{2}+\mu\right)}$ et $\beta_{n}=\frac{1}{(2\pi)^{\frac{n}{2}}}.$\\
\\
{\bf TH\'EOR\`EME 2}. Pour $0<\mu<\frac{1}{2}$ et $-\frac{n}{2}<q<-\frac{\mu}{2}-\frac{n}{4}$, le probl\`eme de Cauchy $(E_{1}),~(C_{2})$ admet la solution unique donn\'ee par:
$$U(t,x)=\int_{R^{n}}f(x')N_{-\mu}(t,x,x')dx'+\frac{t^{2\mu}}{2\mu}\int_{R^{n}}g(x')N_{\mu}(t,x,x')dx'$$
o\`u  $N_{\mu}(t,x,x')=\\
\left\{
\begin{array}{rr}
\frac{2^{2q+\frac{n}{2}}\i^{2q}\Gamma(q+\frac{n}{2})}{(2\pi)^{n}\Gamma(-q)}\left|x-x'\right|^{-2q-n}~_{2}F_{1}(q+\frac{n}{2},q+1,\mu+1,\frac{t^{2}}{\left|x-x'\right|^{2}})~\mbox{si}~0<t<\left|x-x'\right|\\
\frac{2^{2q+\frac{n}{2}}\i^{2q}\Gamma(1+\mu)\Gamma(q+\frac{n}{2})}{(2\pi)^{n}\Gamma(\frac{n}{2})\Gamma(1+\mu-q-\frac{n}{2})}t^{-2q-n}~_{2}F_{1}(q+\frac{n}{2},q+\frac{n}{2}-\mu,\frac{n}{2},\frac{\left|x-x'\right|^{2}}{t^{2}})~\mbox{si}~\left|x-x'\right|<t\\
\end{array}\right.$\\
\\
avec $_{2}F_{1}(a,b,c,z)=\sum^{\infty}_{n=0}\frac{(a)_{n}(b)_{n}}{(c)_{n}n!}z^{n}$ la fonction hyperg\'eom\'etrique de Gauss\\
et $(a)_{n}=\frac{\Gamma(a+n)}{\Gamma(a)}$ le symbole de Pochhammer.\\
\\
{\bf TH\'EOR\`EME 3}. Pour $\nu > -\frac{1}{2}$ et $0<\mu<\frac{1}{2}$, le probl\`eme de Cauchy $(E_{2}),~(C_{1})$ admet la solution unique donn\'ee par:
$$U(t,x)=\int^{+\infty}_{0}f(x')t^{2\mu}K_{-\mu}(t,x,x')x'^{1-2\nu}dx'+\frac{1}{2\mu}\int^{+\infty}_{0}g(x')K_{\mu}(t,x,x')x'^{1-2\nu}dx'$$
o\`u  $K_{\mu}(t,x,x')=\\
\left\{
\begin{array}{rrrrr}
0~~~\mbox{pour}~0<x'<x-t~\mbox{ou}~x'>x+t,\\
\frac{2^{\mu-\frac{1}{2}}\Gamma(1+\mu)}{\sqrt{\pi}\Gamma(\frac{1}{2}+\mu)}(xx')^{\nu+\mu-1}(1-z)^{\mu-\frac{1}{2}}~_{2}F_{1}(\frac{1}{2}-\nu,\frac{1}{2}+\nu,\frac{1}{2}+\mu,\frac{1-z}{2})\\
\mbox{pour}~\left|x-t\right|<x'<x+t,\\
\frac{2^{\mu-\nu}\Gamma(1+\mu)\Gamma(1-\mu+\nu)\sin[(\mu-\nu)\pi]}{\pi\Gamma(\nu+1)}(xx')^{\nu+\mu-1}z^{\mu-\nu-1}\\
\times_{2}F_{1}(\frac{\nu-\mu+1}{2},\frac{\nu-\mu}{2}+1,\nu+1,\frac{1}{z^{2}})~\mbox{pour}~0<x'<t-x\\
\end{array}\right.$ avec $z=\frac{x^{2}+x'^{2}-t^{2}}{2xx'}.$\\
\\
{\bf TH\'EOR\`EME 3 bis}. Pour $0<\mu<1$ et les donn\'ees initiales analytiques\\
$f(x)=\sum^{\infty}_{l=0}a_{l}x^{l}$ et $g(x)=\sum^{\infty}_{l=0}b_{l}x^{l}$,\\
le probl\`eme $(E_{2}),~(C_{1})$ admet la solution unique donn\'ee par:
$$U(t,x)=\sum^{\infty}_{l=0}a_{l}U_{l}+\sum^{\infty}_{l=0}b_{l}V_{l}$$
o\`u $U_{l}=x^{l}~_{2}F_{1}(-\frac{l}{2},\nu-\frac{l}{2},1-\mu,\frac{t^{2}}{x^{2}})$ et $V_{l}=\frac{t^{2\mu}}{2\mu}x^{l}~_{2}F_{1}(-\frac{l}{2},\nu-\frac{l}{2},1+\mu,\frac{t^{2}}{x^{2}})$.\\
\\
{\bf TH\'EOR\`EME 4}. Pour $\nu>-\frac{1}{2},~0<\mu<\frac{1}{2}$ et $-\frac{1}{2}<q<-\frac{\mu}{2}-\frac{1}{4}$, le probl\`eme de Cauchy $(E_{2}),~(C_{2})$ admet la solution unique donn\'ee par:\\
\\ 
$U(t,x)=K(\nu,q)\int^{+\infty}_{0}f(x')H_{-\mu}(t,x,x')x'^{1-2\nu}dx'\\
+\frac{K(\nu,q)t^{2\mu}}{2\mu}\int^{+\infty}_{0}g(x')H_{\mu}(t,x,x')x'^{1-2\nu}dx'$ avec 
$$H_{\mu}(t,x,x')=x^{2\nu}x'^{-2(q+1)}F_{4}(q+1,q+1+\nu,1+\mu,1+\nu,\frac{t^{2}}{x'^{2}},\frac{x^{2}}{x'^{2}}),$$ 
$K(\nu,q)=\frac{2^{2q+1}\i^{2q}\Gamma(q+1+\nu)}{\Gamma(1+\nu)\Gamma(-q)}$\\
et $F_{4}$ la fonction hyperg\'eom\'etrique de deux variables d'Appell d\'efinie par[8]:
$$F_{4}(a,b,c,d,x,y)=\sum^{\infty}_{m,n=0}\frac{(a)_{m+n}(b)_{m+n}}{(c)_{m}(d)_{n}m!n!}x^{m}y^{n}.$$ 
{\bf 2. PR\'ELIMINAIRES}.\\ 
Rappelons l'\'equation de Bessel $[6]~P106$\\
$[\frac{\partial^{2}}{\partial x^{2}}+\frac{1-2\alpha}{x}\frac{\partial}{\partial x}+(\beta\gamma x^{\gamma-1})^{2}+\frac{\alpha^{2}-\nu^{2}\gamma^{2}}{x^{2}}]V=0$\\
dont deux solutions ind\'ependantes sont $t^{\alpha}J_{\nu}(\beta t^{\gamma})$ et $t^{\alpha}Y_{\nu}(\beta t^{\gamma})$ avec $J_{\nu}$ et $Y_{\nu}$ des fonctions de Bessel du premier esp\`ece. On d\'efinit la transformation de Fourier-Bessel-Hankel d'une fonction $f$ par:
$$\widehat{f}(\lambda)=\int^{+\infty}_{0}f(x)(\lambda x)^{\nu}J_{\nu}(\lambda x)x^{1-2\nu}dx.$$
Dans la suite, on aura besoin des lemmes suivants:\\ 
{\bf Lemme 1}. $([6]~P~134-135)$ Pour $\mu>0$ on a les comportements asymptotiques\\
{\bf i} $J_{\mu}(Z)\approx\frac{Z^{\mu}}{2^{\mu}\Gamma(\mu+1)}$ et $Y_{\mu}(Z)\approx\frac{-2^{\mu}\Gamma(\mu)}{\pi.Z^{\mu}}$ en z\'ero,\\
{\bf ii} $J_{\mu}(Z)\approx\sqrt{\frac{2}{\pi Z}}\cos(Z-\frac{1}{2}\mu\pi-\frac{1}{4}\pi)$ et $Y_{\mu}(Z)\approx\sqrt{\frac{2}{\pi Z}}\sin(Z-\frac{1}{2}\mu\pi-\frac{1}{4}\pi)$ \`a l'infini.\\
{\bf Lemme 2}.\\ 
{\bf i} $\widehat{\Lambda_{x}f}(\lambda)=-\lambda^{2}\widehat{f}(\lambda)$.\\
{\bf ii} La transformation inverse de Fourier-Bessel-Hankel est donn\'ee par:\\ 
$f(x)= \int^{+\infty}_{0}\widehat{f}(\lambda)(\lambda x)^{\nu}J_{\nu}(\lambda x)\lambda^{1-2\nu}d\lambda.$\\
{\bf Lemme 3} $([5]~P~675)$.\\ 
{\bf i} La transformation de Fourier d'une fonction radiale est donn\'ee par:\\
$\int_{I\!R^{n}}f(\left|\xi\right|)\exp(i\xi.x)d\xi=\left|x\right|^{1-\frac{n}{2}}\int^{+\infty}_{0}f(r)J_{\frac{n}{2}-1}(r\left|x\right|)r^{\frac{n}{2}}dr.$\\
{\bf ii}. $\int^{+\infty}_{0}r^{-\rho}J_{\mu}(ar)J_{\nu}(br)dr=\frac{2^{-\rho}a^{\rho-\nu-1}b^{\nu}\Gamma(\frac{1+\nu+\mu-\rho}{2})}{\Gamma(1+\nu)\Gamma(\frac{1-\nu+\mu+\rho}{2})}\\
\times~_{2}F_{1}(\frac{1+\nu+\mu-\rho}{2},\frac{1+\nu-\mu-\rho}{2},\nu+1,\frac{b^{2}}{a^{2}})$ o\`u $\nu+\mu-\rho+1>0,~\rho>-1,~a>b>0.$\\
{\bf Lemme 4} $([5]~P~677)$.\\
$\int^{+\infty}_{0}\lambda^{2a-1-\mu}J_{\mu}(\lambda t)J_{\nu}(\lambda x)J_{\nu}(\lambda x')d\lambda=\\ 
\frac{2^{2a-1-\mu}\Gamma(a+\nu)}{\Gamma(1+\mu)\Gamma(1+\nu)\Gamma(1-a)}t^{\mu}x^{\nu}x'^{-\nu-2a}.F_{4}(a,a+\nu,1+\mu,1+\nu,\frac{t^{2}}{x'^{2}},\frac{x^{2}}{x'^{2}})$\\
pour $-\nu<a<\frac{5}{4}+\frac{\mu}{2},~x>0,~t>0$ et $x'>x+t$;\\ 
et pour $-\nu<a<\frac{3}{4}+\frac{\mu}{2}$ l'integrale converge absolument\\  
et par suite, elle prolonge $F_{4}$ pour $0<x'<x+t$.\\  
{\bf Lemme 5}. Pour $W_{n,\mu}(t,x,x')=C_{n,\mu}\left[t^{2}-\left|x'-x\right|^{2}\right]^{\mu-\frac{n}{2}}$ et $C_{n,\mu}=\frac{\Gamma(1+\mu)}{\pi^{\frac{n}{2}}\Gamma(1+\mu-\frac{n}{2})}$, on a\\
{\bf i} $W_{n,\mu}(t,x,x')=\left\{
\begin{array}{rr}
\alpha_{n,\mu}(\frac{\partial}{t\partial t})^{\frac{n-1}{2}}\left[t^{2}-\left|x'-x\right|^{2}\right]^{\mu-\frac{1}{2}}~\mbox{si n est impair}\\ 
\beta_{n}(\frac{\partial}{t\partial t})^{\frac{n}{2}}\left[t^{2}-\left|x'-x\right|^{2}\right]^{\mu}~\mbox{si n est pair}\\
\end{array}\right.$\\
\\
{\bf ii} $\Delta_{n}W_{\mu}(t,x,x')=2(\mu-\frac{n}{2})C_{n,\mu}(t^{2}-\left|x-x'\right|^{2})^{\mu-\frac{n}{2}-2}[2(\mu-1)\left|x-x'\right|^{2}-nt^{2}]$,\\
\\
{\bf iii} $[\frac{\partial^{2}}{\partial t^{2}}+ \frac{1-2\mu}{t}\frac{\partial}{\partial t}]W_{\mu}(t,x,x')=\\
4(\mu-\frac{n}{2})C_{n,\mu}(t^{2}-\left|x-x'\right|^{2})^{\mu-\frac{n}{2}-2}[(1-\mu)(t^{2}-\left|x-x'\right|^{2})+(\mu-\frac{n}{2}-1)t^{2}]$,\\
{\bf Preuve.} Une simple v\'erification suffit.\\
\\
{\bf 3. \'EQUATION CLASSIQUE D'EULER-POISSON-DARBOUX}.\\
{\bf Preuve du th\'eor\`eme 1.} D'apr\`es le lemme 5 on a\\
$U(t,x)=t^{2\mu}\int_{\left|x'-x\right|< t}f(x')W_{-\mu}(t,x,x')dx'+\frac{1}{2\mu}\int_{\left|x'-x\right| <t}g(x')W_{\mu}(t,x,x')dx'$,\\
et $(\frac{\partial^{2}}{\partial t^{2}}+ \frac{1-2\mu}{t}\frac{\partial}{\partial t}-\Delta_{n})W_{\mu}=0$,\\ 
soit $V$ une solution de l'\'equation $(E_{1})$, remarquons que si $V(t,x)=t^{2\mu}w_{-\mu}(t,x)$ alors
$$\Delta w_{-\mu}(t,x)=[\frac{\partial^{2}}{\partial t^{2}}+ \frac{1+2\mu}{t}\frac{\partial}{\partial t}]w_{-\mu}(t,x),$$ 
donc il suffit de montrer que  $W_{\mu}$ v\'erifie l'\'equation $(E_{1})$, ce qui est r\'ealis\'e.\\
\\
Pour voir les conditions initiales, on utilise les coordonn\'ees polaires centr\'ees en $x$\\ 
$x'=x+r\omega,~\omega\in S^{n-1},~S^{n-1}=\left\{\omega\in R^{n},~ \left|\omega\right|=1\right\}$ et le changement des variables $r=ts,~0 <s <1$, on obtient $$U\left(t,x\right)=C_{n,-\mu}\int^{1}_{0}f^{\#}_{x}(ts)(1-s^{2})^{-\mu-\frac{n}{2}}s^{n-1}ds+\frac{C_{n,\mu}}{2\mu}t^{2\mu}\int^{1}_{0}g^{\#}_{x}(ts)(1-s^{2})^{\mu-\frac{n}{2}}s^{n-1}ds$$
avec $f^{\#}_{x}(r)=\int_{S^{n-1}}f(x+r\omega)d\sigma(\omega)$,\\
et $\int_{S^{n-1}}d\sigma(\omega)=\frac{2\pi^{\frac{n}{2}}}{\Gamma\left(\frac{n}{2}\right)} $.\\
\`a la limite on obtient la premi\`ere donn\'ee initiale \`a savoir que
$$\int^{1}_{0}(1-s^{2})^{-\mu-\frac{n}{2}}s^{n-1}ds=\frac{1}{2}B(1-\mu-\frac{n}{2},\frac{n}{2})=\frac{\Gamma(1-\mu-\frac{n}{2})\Gamma(\frac{n}{2})}{2\Gamma(1-\mu)}.$$
De m\^eme on obtient la deuxi\`eme donn\'ee initiale.\\
{\bf Preuve du th\'eor\`eme 2.} On pose $F(x)=\Delta^{q}_{x}f(x)$ et $G(x)=\Delta^{q}_{x}g(x)$, en utilisant la transformation de Fourier et les lemmes 1, 2 et quelques propri\'et\'es des fonctions $J_{\nu}$ et $Y_{\nu}$ (voir $[6]$ et $[7]$) on obtient
$$\widehat{U}(t,\xi)\approx\frac{\left|\xi\right|^{\mu}C_{1}(\xi)}{2^{\mu}\Gamma(\mu+1)}t^{2\mu}-\frac{2^{\mu}\Gamma(\mu)}{\pi\left|\xi\right|^{\mu}}C_{2}(\xi)\Rightarrow \widehat{U}(0,\xi)=-\frac{2^{\mu}\Gamma(\mu)C_{2}(\xi)}{\pi\left|\xi\right|^{\mu}},$$
on obtient $C_{2}(\xi)=-\frac{\pi\left|\xi\right|^{\mu}\widehat{F}(\xi)}{2^{\mu}\Gamma(\mu)}$, soit $Z=\left|\xi\right| t$ on a
$$\frac{\partial}{\partial t}\widehat{U}(t,\xi)=
\left|\xi\right|^{1-\mu}C_{1}(\xi)Z^{\mu}J_{\mu-1}(Z)-\frac{\pi\left|\xi\right|\widehat{F}(\xi)}{2^{\mu}\Gamma(\mu)}Z^{\mu}Y_{\mu-1}(Z)$$
$=\left|\xi\right|^{1-\mu}C_{1}(\xi)Z^{\mu}\left\{\cos[(1-\mu)\pi]J_{1-\mu}(Z)-\sin[(1-\mu)\pi]Y_{1-\mu}(Z)\right\}\\
-\frac{\pi\left|\xi\right|\widehat{F}(\xi)}{2^{\mu}\Gamma(\mu)}Z^{\mu}\left\{\sin[(1-\mu)\pi]J_{1-\mu}(Z)+\cos[(1-\mu)\pi]Y_{1-\mu}(Z)\right\}$,\\
\\
$\frac{\partial}{\partial t}\widehat{U}(t,\xi)\approx \frac{t}{2^{1-\mu}\Gamma(2-\mu)}\left\{\cos[(1-\mu)\pi]\left|\xi\right|^{2-\mu}C_{1}(\xi)-\frac{\pi\sin[(1-\mu)\pi]}{2^{\mu}\Gamma(\mu)}\left|\xi\right|^{2}\widehat{F}(\xi)\right\}\\
~~~~~~~~~~~~~~~~+\frac{2^{1-\mu}\Gamma(1-\mu)}{\pi}t^{2\mu-1}\left\{\sin[(1-\mu)\pi]\left|\xi\right|^{\mu}C_{1}(\xi)+\frac{\pi \cos[(1-\mu)\pi]}{2^{\mu}\Gamma(\mu)}\left|\xi\right|^{2\mu}\widehat{F}(\xi)\right\}$,\\
\\
$\lim_{t\rightarrow 0} t^{1-2\mu}\frac{\partial}{\partial t}\widehat{U}(t,\xi)=\\ \frac{2^{1-\mu}\Gamma(1-\mu)}{\pi}\left\{\sin[(1-\mu)\pi]\left|\xi\right|^{\mu}C_{1}(\xi) 
+\frac{\pi \cos[(1-\mu)\pi]}{2^{\mu}\Gamma(\mu)}\left|\xi\right|^{2\mu}\widehat{F}(\xi)\right\}=\widehat{G}(\xi)$\\
\\
$\Rightarrow C_{1}(\xi)= -\frac{\pi\left|\xi\right|^{\mu}\widehat{F}(\xi)}{2^{\mu}\tan[(1-\mu)\pi]\Gamma(\mu)} +\frac{\pi\left|\xi\right|^{-\mu}\widehat{G}(\xi)}{2^{1-\mu}\sin[(1-\mu)\pi]\Gamma(1-\mu)}~~~~$, par suite on a\\
\\
$\widehat{U}(t,\xi)=\\
\\
\left\{-\frac{\pi\left|\xi\right|^{\mu}\widehat{F}(\xi)}{2^{\mu}\tan[(1-\mu)\pi]\Gamma(\mu)} +\frac{\pi\left|\xi\right|^{-\mu}\widehat{G}(\xi)}{2^{1-\mu}\sin[(1-\mu)\pi]\Gamma(1-\mu)}\right\}t^{\mu}J_{\mu}(\left|\xi\right| t) -\frac{\pi\left|\xi\right|^{\mu}\widehat{F}(\xi)}{2^{\mu}\Gamma(\mu)}t^{\mu}Y_{\mu}(\left|\xi\right| t)\\
\\
=\frac{\pi\left|\xi\right|^{-\mu}\widehat{G}(\xi)}{2^{1-\mu}\sin[(1-\mu)\pi]\Gamma(1-\mu)}t^{\mu}J_{\mu}(\left|\xi\right| t)
-\frac{\pi\left|\xi\right|^{\mu}\widehat{F}(\xi)}{2^{\mu}\Gamma(\mu)}t^{\mu}\left\{\frac{1}{\tan[(1-\mu)\pi]}J_{\mu}(\left|\xi\right| t)+ Y_{\mu}(\left|\xi\right| t)\right\}$, or
$$\frac{1}{\tan[(1-\mu)\pi]}J_{\mu}(\left|\xi\right| t)+ Y_{\mu}(\left|\xi\right| t)=-\frac{J_{-\mu}(\left|\xi\right| t)}{\sin[(1-\mu)\pi]}~\mbox{et}~\Gamma(\mu)\Gamma(1-\mu)=\frac{\pi}{\sin[(1-\mu)\pi]},$$
alors $\widehat{U}(t,\xi)=2^{-\mu}\Gamma(1-\mu)t^{\mu}\left|\xi\right|^{\mu}J_{-\mu}(\left|\xi\right| t)\widehat{F}(\xi)+2^{\mu-1}\Gamma(\mu)t^{\mu}\left|\xi\right|^{-\mu}J_{\mu}(\left|\xi\right| t)\widehat{G}(\xi)$, donc
$$\widehat{U}(t,\xi)=2^{-\mu}\i^{2q}\Gamma(1-\mu)t^{\mu}\left|\xi\right|^{2q+\mu}J_{-\mu}(\left|\xi\right| t)\widehat{f}(\xi)+2^{\mu-1}\i^{2q}\Gamma(\mu)t^{\mu}\left|\xi\right|^{2q-\mu}J_{\mu}(\left|\xi\right| t)\widehat{g}(\xi).$$
La transformation inverse de Fourier, l'interversion des int\'egrales et le lemme 3 nous donnent le r\'esultat du th\'eor\`eme 2.\\ 
{\bf Remarque}. On justifie l'interversion des int\'egrales \`a l'aide de Fubini, car les int\'egrales qui repr\'esentent les noyaux convergent absolument\\ 
(voir le lemme 3 et les comportements asymptotiques des fonctions de Bessel).\\
\\
{\bf 4. \'EQUATION RADIALE D'EULER-POISSON-DARBOUX}.\\
{\bf - Preuve du th\'eor\`eme 3.}\\
Soit $\varphi$ une solution de l'\'equation $(E_{1})$, remarquons que si $\varphi(t,x)=t^{2\mu}k_{-\mu}(t,x)$ alors
$$[\frac{\partial^{2}}{\partial x^{2}}+\frac{1-2\nu}{x}\frac{\partial}{\partial x}]k_{-\mu}(t,x)=[\frac{\partial^{2}}{\partial t^{2}}+ \frac{1+2\mu}{t}\frac{\partial}{\partial t}]k_{-\mu}(t,x)~(4.1),$$ 
donc il suffit de montrer que le noyaux $K_{\mu}(t,x,x')$ v\'erifie l'\'equation $(E_{1})$, pour cela on fait le changement des fonctions
$\varphi(t,x)=x^{\nu+\mu-1}\psi(t,x)$, on obtient
$$[\frac{\partial^{2}}{\partial x^{2}}+\frac{2\mu-1}{x}\frac{\partial}{\partial x}+\frac{(\mu-1)^{2}-\nu^{2}}{x^{2}}]\psi(t,x)= [\frac{\partial^{2}}{\partial t^{2}}+ \frac{1-2\mu}{t}\frac{\partial}{\partial t}]\psi(t,x)~~(4.2),$$ 
on pose $\psi(t,x)=F(z)$ avec $z=\frac{x^{2}+x'^{2}-t^{2}}{2xx'}$ alors
$$[(1-z^{2})\frac{\partial^{2}}{\partial z^{2}}+(2\mu-3)z\frac{\partial}{\partial z}+\nu^{2}-(\mu-1)^{2}]F(z)=0~~(4.3),$$ 
finalement, pour $F(z)=(1-z^{2})^{\frac{\mu}{2}-\frac{1}{4}}G(z)$ on obtient\\ l'\'equation de Legendre $[7]~P~198$
$$[(1-z^{2})\frac{\partial^{2}}{\partial z^{2}}-2z\frac{\partial}{\partial z}+(\nu^{2}-\frac{1}{4})-\frac{(\frac{1}{2}-\mu)^{2}}{1-z^{2}}]G(z)=0~(4.4),$$
dont deux solutions sont $P^{\frac{1}{2}-\mu}_{\nu-\frac{1}{2}}(z)$ et $Q^{\frac{1}{2}-\mu}_{\nu-\frac{1}{2}}(z)$ o\`u\\ $P^{\mu}_{\nu}(z)=\frac{1}{\Gamma(1-\mu)}(\frac{1+z}{1-z})^{\frac{\mu}{2}}~_{2}F_{1}(-\nu,\nu+1,1-\mu,\frac{1-z}{2})~\mbox{pour}~\left|z-1\right|<2$, et\\
\\
$Q^{\mu}_{\nu}(z)=e^{i\pi\mu}\frac{\sqrt{\pi}\Gamma(\nu+\mu+1)}{2^{\nu+1}\Gamma(\nu+\frac{3}{2})}(z^{2}-1)^{\frac{\mu}{2}}z^{-\nu-\mu-1}~_{2}F_{1}(\frac{\nu+\mu}{2}+1,\frac{\nu+\mu+1}{2},\nu+\frac{3}{2},\frac{1}{z^{2}})$\\
lorsque $\left|z\right|>1$.\\
Pour les conditions initiales on prend $t<x$, on obtient\\
$U(t,x)=\\
\frac{2^{-2\mu-1}\Gamma(1-\mu)}{\sqrt{\pi}\Gamma(\frac{1}{2}-\mu)}t^{2\mu}\int^{x+t}_{x-t}f(x')(xx')^{\nu-\mu-1}X^{-\mu-\frac{1}{2}}~_{2}F_{1}(\frac{1}{2}-\nu,\frac{1}{2}+\nu,\frac{1}{2}-\mu,X)x'^{1-2\nu}dx'\\
+\frac{4^{\mu-1}\Gamma(\mu)}{\sqrt{\pi}\Gamma(\frac{1}{2}+\mu)}\int^{x+t}_{x-t}g(x')(xx')^{\nu+\mu-1}X^{\mu-\frac{1}{2}}~_{2}F_{1}(\frac{1}{2}-\nu,\frac{1}{2}+\nu,\frac{1}{2}+\mu,X)x'^{1-2\nu}dx'$,\\
\\
le changement des variables $x'=x+ts$ donne\\
$U(t,x)=\frac{\Gamma(1-\mu)}{\sqrt{\pi}\Gamma(\frac{1}{2}-\mu)}\\
\times\int^{1}_{-1}f(x+ts)x^{\nu-\frac{1}{2}}(x+ts)^{-\nu+\frac{1}{2}}(1-s^{2})^{-\mu-\frac{1}{2}}~_{2}F_{1}(\frac{1}{2}-\nu,\frac{1}{2}+\nu,\frac{1}{2}-\mu,\frac{t^{2}(1-s^{2})}{4x(x+ts)})ds\\
+\frac{\Gamma(\mu)t^{2\mu}}{2\sqrt{\pi}\Gamma(\frac{1}{2}+\mu)}\\
\times\int^{1}_{-1}g(x+ts)x^{\nu-\frac{1}{2}}(x+ts)^{-\nu+\frac{1}{2}}(1-s^{2})^{\mu-\frac{1}{2}}~_{2}F_{1}(\frac{1}{2}-\nu,\frac{1}{2}+\nu,\frac{1}{2}+\mu,\frac{t^{2}(1-s^{2})}{4x(x+ts)})ds$,\\
\\
\`a la limite on obtient la premi\`ere donn\'ee initiale \`a savoir que
$$\int^{1}_{-1}(1-s^{2})^{-\mu-\frac{1}{2}}ds=2^{-2\mu}B(\frac{1}{2}-\mu,\frac{1}{2}+\mu)=\frac{2^{-2\mu}[\Gamma(\frac{1}{2}-\mu)]^{2}}{\Gamma(1-2\mu)}=\frac{\sqrt{\pi}\Gamma(\frac{1}{2}-\mu)}{\Gamma(1-\mu)}.$$
De m\^eme on obtient la deuxi\`eme donn\'ee initiale.\\
\\
{\bf - Preuve du th\'eor\`eme 3 bis.} D'apr\`es le principe de superposition, il suffit d'\'etudier les probl\`emes de Cauchy [2]
$$\Lambda^{\nu}_{x}U_{l}=\Lambda^{\mu}_{t}U_{l},~\Lambda^{\nu}_{x}=\Lambda_{x},~U_{l}(0,x)=x^{l},~\lim_{t\rightarrow0}t^{1-2\mu}\frac{\partial}{\partial t}U\left(t,x\right)=0~~~~(P_{1}).$$
$$\Lambda^{\nu}_{x}V_{l}=\Lambda^{\mu}_{t}V_{l},~V_{l}(0,x)=0,~\lim_{t\rightarrow0}t^{1-2\mu}\frac{\partial}{\partial t}V\left(t,x\right)=x^{l}~~~~(P_{2}).$$
Pour r\'esoudre $(P_{1})$, on pose $U_{l}=x^{l}\phi(Z)$ avec $Z=\frac{t^{2}}{x^{2}}$ et $\left|Z\right|<1$, on obtient l'\'equation
$$Z(1-Z)\frac{\partial^{2}\phi}{\partial Z^{2}}+[1-\mu-(\nu-l+1)Z]\frac{\partial\phi}{\partial Z}+\frac{l}{2}(\nu-\frac{l}{2})\phi=0~~(4.5).$$ 
Or $1-\mu\notin Z$, la solution g\'en\'erale de $(4.5)$ s'\'ecrit sous la forme $[6]P248$\\
$\phi(Z)=A~_{2}F_{1}(-\frac{l}{2},\nu-\frac{l}{2},1-\mu,Z)+BZ^{\mu}~_{2}F_{1}(\mu-\frac{l}{2},\nu+\mu-\frac{l}{2},1+\mu,Z)$.\\
\\
Les conditions initiales pour $U_{l}$ donnent $A=1$ et $B=0$, et par suite on obtient\\
$~~~~~~U_{l}=x^{l}~_{2}F_{1}(-\frac{l}{2},\nu-\frac{l}{2},1-\mu,Z).$
\\
De la m\^eme mani\`ere pour r\'esoudre $(P_{2})$, on pose $V_{l}=\frac{t^{2\mu}}{2\mu}x^{l}\psi(Z)$\\
avec $Z=\frac{t^{2}}{x^{2}}$ et $\left|Z\right|<1$, on obtient l'\'equation
$$Z(1-Z)\frac{\partial^{2}\psi}{\partial Z^{2}}+[1+\mu-(\nu-l+1)Z]\frac{\partial\phi}{\partial Z}+\frac{l}{2}(\nu-\frac{l}{2})\phi=0~~(4.6),$$
$1+\mu\notin Z$, la solution g\'en\'erale de $(4.6)$ s'\'ecrit sous la forme
$$\psi(Z)=A'~_{2}F_{1}(-\frac{l}{2},\nu-\frac{l}{2},1+\mu,Z)+B'Z^{-\mu}~_{2}F_{1}(-\mu-\frac{l}{2},\nu-\mu-\frac{l}{2},1-\mu,Z).$$
Les conditions initiales pour $V_{l}$ donnent $A'=1$ et $B'=0$, et par suite on obtient\\
$V_{l}=\frac{t^{2\mu}}{2\mu}x^{l}~_{2}F_{1}(-\frac{l}{2},\nu-\frac{l}{2},1+\mu,Z).$\\
\\
{\bf - Preuve du th\'eor\`eme 4.} 
par une m\'ethode analogue \`a celle proced\'ee dans la preuve du th\'eor\`eme 2 on obtient
$$\widehat{U}(t,\lambda)=2^{-\mu}\Gamma(1-\mu)t^{\mu}\lambda^{\mu}J_{-\mu}(\lambda t)\widehat{F}(\lambda)+2^{\mu-1}\Gamma(\mu)t^{\mu}\lambda^{-\mu}J_{\mu}(\lambda t)\widehat{G}(\lambda).$$
La transformation inverse de Fourier-Bessel-Hankel, l'interversion des int\'egrales et le lemme 4 nous donnent le r\'esultat du th\'eor\`eme 4.\\ 
\\
{\bf PROPOSITION}.\\
$U(t,x)=x^{\alpha}(x^{2}-t^{2})^{\beta}F_{4}(\frac{-\alpha}{2},\frac{-\alpha}{2}+\nu,1-\mu,\gamma,\frac{t^{2}}{x^{2}},\frac{(x^{2}-t^{2})^{2}}{x^{2}})$\\
v\'erifie l'\'equation $(E_{2})$ avec $\beta=\mu+\nu-\alpha-1$\\ 
\\
{\bf Preuve}. On rappelle d'abord que la fonction $F_{4}(a,b,c,d,x,y)$ v\'erifie le syst\`eme de deux \'equations [8]\\
$\left\{
\begin{array}{rr}
y(1-y)\frac{\partial^{2}}{\partial y^{2}}-z^{2}\frac{\partial^{2}}{\partial z^{2}}-2yz\frac{\partial^{2}}{\partial y \partial z}+[c-(a+b+1)y]\frac{\partial}{\partial y}-(a+b+1)z\frac{\partial}{\partial z}-ab=0~(1)\\
z(1-z)\frac{\partial^{2}}{\partial z^{2}}-y^{2}\frac{\partial^{2}}{\partial y^{2}}-2yz\frac{\partial^{2}}{\partial y \partial z}+[d-(a+b+1)z]\frac{\partial}{\partial z}-(a+b+1)y\frac{\partial}{\partial y}-ab=0~(2).\\
\end{array}\right.$\\
\\
On cherche maintenant une solution de $(E_{2})$ sous la forme \\
\\
$V(t,x)=x^{\alpha}(x^{2}-t^{2})^{\beta}W(t,x)$, on obtient que $W$ v\'erifie l'\'equation\\
\\
$x^{2}\frac{\partial^{2}W}{\partial x^{2}}+(1-2\nu+2\alpha+4\beta\frac{x^{2}}{x^{2}-t^{2}})x\frac{\partial W}{\partial
x}=\\x^{2}\frac{\partial^{2}W}{\partial t^{2}}+(\frac{1-2\mu}{t}-4\beta\frac{t}{x^{2}-t^{2}})x^{2}\frac{\partial W}{\partial
t}+\alpha(2\nu-\alpha)W~~(4.7)$;\\
\\
en posant $W(t,x)=F(y,z)$ avec $y=\frac{t^{2}}{x^{2}}$ et $z=\frac{(x^{2}-t^{2})^{2}}{x^{2}}$ on obtient que $F$ v\'erifie l'\'equation $(1)$ du syst\`eme.\\ 
\\ 
{\bf 5. APPLICATIONS ET PERSPECTIVES}.\\ 
{\bf Corollaire 1} ( \'Equation des ondes en dimension $n$ [4] ).\\ 
Pour $\mu\rightarrow\frac{1}{2}$ dans le th\'eor\`eme 1, on retrouve la solution du probl\`eme de Cauchy pour l'\'equation des ondes classique en dimension $n$\\
\\
$U(t,x)=b(N)\frac{\partial}{\partial t}(\frac{1}{t}\frac{\partial}{\partial t})^{N-1}[t^{2N-1}\int_{\left\{\left|y\right|=1\right\}}\Phi(x-ty)d\sigma(y)]+\\
\\
b(N)(\frac{1}{t}\frac{\partial}{\partial t})^{N-1}[t^{2N-1}\int_{\left\{\left|y\right|=1\right\}}\Psi(x-ty)d\sigma(y)]$ si $n$ est impair $(n=2N+1)$\\
\\
o\`u $b(N)=2^{-1}[1.3.5...(2N-1)]^{-1}\pi^{-N-\frac{1}{2}}\Gamma(N+\frac{1}{2})=\frac{1}{2(2\pi)^{N}}$\\
et $d\sigma(y)$ est la mesure de surface $\left\{\left|y\right|=1\right\}$ ,\\
\\
$U(t,x)=2b(N)\frac{\partial}{\partial t}(\frac{1}{t}\frac{\partial}{\partial t})^{N-1}[t^{2N-1}\int_{\left\{\left|y\right|<1\right\}}\frac{\Phi(x-ty)}{\sqrt{1-\left|y\right|^{2}}}dy]+\\
\\
2b(N)(\frac{1}{t}\frac{\partial}{\partial t})^{N-1}[t^{2N-1}\int_{\left\{\left|y\right|<1\right\}}\frac{\Psi(x-ty)}{\sqrt{1-\left|y\right|^{2}}}dy]$ si $n$ est pair $(n=2N)$.\\
{\bf Preuve}. On distingue deux cas:\\
{\bf - Cas $n$ impair $(n=2N+1)$.} $U(t,x)=I_{1}(t,x)+J_{1}(t,x),\\
I_{1}(t,x)=\frac{\Gamma(\frac{1}{2})}{2^{N}\pi^{N+\frac{1}{2}}}\lim_{\mu\rightarrow\frac{1}{2}}\frac{1}{\Gamma(\frac{1}{2}-\mu)}t(\frac{\partial}{t\partial t})^{N}\int_{\left|x'-x\right|<t}f\left(x'\right)\left(t^{2}-\left|x'-x\right|^{2}\right)^{-\mu-\frac{1}{2}}dx',\\
et~J_{1}(t,x)=\frac{1}{2(2\pi)^{N}}\left(\frac{\partial}{t\partial t}\right)^{N}\int_{\left|x'-x\right|< t}g\left(x'\right)dx'$,\\
$I_{1}(t,x)=\frac{1}{2(2\pi)^{N}}\lim_{\mu\rightarrow\frac{1}{2}}t(\frac{\partial}{t\partial t})^{N}\frac{1}{t}\frac{\partial}{\partial t}\int_{\left|x'-x\right|< t}f\left(x'\right)\left(t^{2}-\left|x'-x\right|^{2}\right)^{\frac{1}{2}-\mu}dx'\\
=\frac{1}{2(2\pi)^{N}}\frac{\partial}{\partial t}(\frac{\partial}{t\partial t})^{N}\int_{\left|x'-x\right|< t}f\left(x'\right)dx'$\\
\\
et$(\frac{\partial}{t\partial t})^{N}\int_{\left|x'-x\right|< t}f\left(x'\right)dx'=(\frac{\partial}{t\partial t})^{N-1}\frac{1}{t}\frac{\partial}{\partial t}\left\{t^{2N+1}\int^{1}_{0}[\int_{\left|y\right|=1 }f\left(x-tsy\right)d\sigma(y)]s^{2N}ds\right\}$\\
\\
et$\frac{1}{t}\frac{\partial}{\partial t}\left\{t^{2N+1}\int^{1}_{0}[\int_{\left|y\right|=1 }f\left(x-tsy\right)d\sigma(y)]s^{2N}ds\right\}=\\(2N+1)t^{2N-1}\int^{1}_{0}[\int_{\left|y\right|=1 }f\left(x-tsy\right)d\sigma(y)]s^{2N}ds\\-t^{2N}\int^{1}_{0}[\int_{\left|y\right|=1 }f'\left(x-tsy\right)yd\sigma(y)]s^{2N+1}ds$\\
\\
et$(2N+1)t^{2N-1}\int^{1}_{0}[\int_{\left|y\right|=1 }f\left(x-tsy\right)d\sigma(y)]s^{2N}ds=\\
t^{2N-1}\left\{[s^{2N+1}\int_{\left|y\right|=1 }f\left(x-tsy\right)d\sigma(y)]^{1}_{0}+t\int^{1}_{0}[\int_{\left|y\right|=1 }f'\left(x-tsy\right)yd\sigma(y)]s^{2N+1}ds\right\}$,\\
\\
donc $I_{1}(t,x)=\frac{1}{2(2\pi)^{N}}\frac{\partial}{\partial t}(\frac{\partial}{t\partial t})^{N-1}\left\{t^{2N-1}\int_{\left|y\right|=1 }f\left(x-ty\right)d\sigma(y)\right\},\\
et~J_{1}(t,x)=\frac{1}{2(2\pi)^{N}}(\frac{\partial}{t\partial t})^{N-1}\left\{t^{2N-1}\int_{\left|y\right|=1 }g\left(x-ty\right)d\sigma(y)\right\}$.\\
\\
{\bf - Cas $n$ pair $(n=2N)$.} $U(t,x)=I_{2}(t,x)+J_{2}(t,x),\\
I_{2}(t,x)=\frac{1}{(2\pi)^{N}}t(\frac{\partial}{t\partial t})^{N}\int_{\left|x'-x\right|< 
t}f\left(x'\right)\left(t^{2}-\left|x'-x\right|^{2}\right)^{-\frac{1}{2}}dx',\\
et~J_{2}(t,x)=\frac{1}{(2\pi)^{N}}(\frac{\partial}{t\partial t})^{N}\int_{\left|x'-x\right|< t}g\left(x'\right)\left(t^{2}-\left|x'-x\right|^{2}\right)^{\frac{1}{2}}dx'$,\\
\\
$I_{2}(t,x)=\frac{1}{(2\pi)^{N}}t(\frac{\partial}{t\partial t})^{N}\frac{1}{t}\frac{\partial}{\partial t}\int_{\left|x'-x\right|< t}f\left(x'\right)\left(t^{2}-\left|x'-x\right|^{2}\right)^{\frac{1}{2}}dx'\\
=\frac{1}{(2\pi)^{N}}\frac{\partial}{\partial t}(\frac{\partial}{t\partial t})^{N}\int_{\left|x'-x\right|< t}f\left(x'\right)\left(t^{2}-\left|x'-x\right|^{2}\right)^{\frac{1}{2}}dx'$\\
\\
et $(\frac{\partial}{t\partial t})^{N}\int_{\left|x'-x\right|< t}f\left(x'\right)\left(t^{2}-\left|x'-x\right|^{2}\right)^{\frac{1}{2}}dx'=\\(\frac{\partial}{t\partial t})^{N-1}\int_{\left|x'-x\right|< t}f\left(x'\right)\left(t^{2}-\left|x'-x\right|^{2}\right)^{-\frac{1}{2}}dx'$,\\
\\
donc $I_{2}(t,x)=\frac{1}{(2\pi)^{N}}\frac{\partial}{\partial t}(\frac{\partial}{t\partial t})^{N-1}\left\{t^{2N-1}\int_{\left|y\right|< 1}f\left(x-ty\right)\left(1-\left|y\right|^{2}\right)^{-\frac{1}{2}}dy\right\}$\\
et $J_{2}(t,x)=\frac{1}{(2\pi)^{N}}(\frac{\partial}{t\partial t})^{N-1}\left\{t^{2N-1}\int_{\left|y\right|< 1}g\left(x-ty\right)\left(1-\left|y\right|^{2}\right)^{-\frac{1}{2}}dy\right\}$.\\ 
\\
{\bf Corollaire 2} ( Th\'eor\`eme 1.1 [4] ).\\
Pour $\nu=-\alpha$ et $\mu\rightarrow\frac{1}{2}$ dans le th\'eor\`eme 3, on retrouve la solution du probl\`eme de Cauchy pour l'\'equation radiale des ondes\\
\\
$U(t,x)=\int^{+\infty}_{0}g(x')K(t,x,x')dx'+\int^{+\infty}_{0}f(x')\frac{\partial}{\partial t}K(t,x,x')dx'\\ +\left\{
\begin{array}{rr}
\frac{1}{2}x^{-\alpha-\frac{1}{2}}[f(x-t)(x-t)^{\frac{1}{2}+\alpha}+f(x+t)(x+t)^{\frac{1}{2}+\alpha}]~\mbox{pour}~t<x\\
\frac{1}{2}x^{-\alpha-\frac{1}{2}}[-\sin\pi\alpha.f(t-x)(t-x)^{\frac{1}{2}+\alpha}+f(t+x)(t+x)^{\frac{1}{2}+\alpha}]~\mbox{pour}~x<t\\
\end{array}\right.$ \\
\\
o\`u $K(t,x,x')=K_{\frac{1}{2}}(t,x,x')x'^{1+2\alpha}=\\
\left\{
\begin{array}{rrrr}
0~~~\mbox{pour}~0<x'<x-t~\mbox{ou}~x'>x+t,\\
\frac{1}{2}x^{-\alpha-\frac{1}{2}}x'^{\frac{1}{2}+\alpha}~_{2}F_{1}(\frac{1}{2}-\alpha,\frac{1}{2}+\alpha,1,\frac{t^{2}-(x'-x)^{2}}{4xx'})~\mbox{pour}~\left|x-t\right|<x'<x+t,\\
\frac{2^{-2\alpha-1}\sqrt{\pi}}{\Gamma(\frac{1}{2}-\alpha)\Gamma(\alpha+1)}x^{-\alpha-\frac{1}{2}}x'^{\alpha+\frac{1}{2}}(\frac{4xx'}{t^{2}-(x'-x)^{2}})^{\alpha+\frac{1}{2}}\\
\times{2}F_{1}(\alpha+\frac{1}{2},\alpha+\frac{1}{2},2\alpha+1,\frac{4xx'}{t^{2}-(x'-x)^{2}})~\mbox{pour}~0<x'<t-x.\\
\end{array}\right.$\\
{\bf Preuve}. D'apr\`es les relations $[7] P41$\\
\\
$\frac{d}{dX}[X^{c-1}~_{2}F_{1}(a,b,c,X)]=(c-1)X^{c-2}~_{2}F_{1}(a+1,b,c-1,X)$\\
et $\frac{d}{dY}[Y^{a}~_{2}F_{1}(a,b,c,Y)]=aY^{a-1}~_{2}F_{1}(a+1,b,c,Y)~~~~~~$, on obtient\\
\\
$U(t,x)=\lim_{\mu\rightarrow\frac{1}{2}}\frac{4^{-\mu-\frac{1}{2}}\Gamma(1-\mu)}{\sqrt{\pi}\Gamma(\frac{3}{2}-\mu)}t^{2\mu}\\
\times\int^{x+t}_{\left|x-t\right|}f(x')(xx')^{-\alpha-\mu-1}\frac{d}{dX}[X^{\frac{1}{2}-\mu}~_{2}F_{1}(\frac{1}{2}-\alpha,\frac{1}{2}+\alpha,\frac{3}{2}-\mu,X)]x'^{1+2\alpha}dx'\\
+[\int^{t-x}_{0}f(x')\frac{\partial}{\partial t}K(t,x,x')dx'~si~t>x]\\ +\int^{+\infty}_{0}g(x')K(t,x,x')dx',~X=\frac{1-z}{2}$\\ 
\\
on distingue deux cas:\\
{\bf - Pour t $<$ x,} on obtient \\
\\
$U(t,x)=\lim_{\mu\rightarrow\frac{1}{2}}\frac{4^{-\mu-\frac{1}{2}}\Gamma(1-\mu)}{\sqrt{\pi}\Gamma(\frac{3}{2}-\mu)}t^{2\mu}x^{-\alpha-\mu-1}\\
\times\int^{x+t}_{x-t}f(x')x'^{\alpha-\mu}X^{\frac{1}{2}-\mu}\frac{d}{dX}~_{2}F_{1}(\frac{1}{2}-\alpha,\frac{1}{2}+\alpha,\frac{3}{2}-\mu,X)dx'\\
+\lim_{\mu\rightarrow\frac{1}{2}}\frac{4^{-\mu-\frac{1}{2}}\Gamma(1-\mu)(\frac{1}{2}-\mu)}{\sqrt{\pi}\Gamma(\frac{3}{2}-\mu)}t^{2\mu}x^{-\alpha-\mu-1}\\ \times\int^{x+t}_{x-t}f(x')x'^{\alpha-\mu}X^{-\frac{1}{2}-\mu}~_{2}F_{1}(\frac{1}{2}-\alpha,\frac{1}{2}+\alpha,\frac{3}{2}-\mu,X)dx'+\int^{+\infty}_{0}g(x')K(t,x,x')dx'$,\\
alors\\
$~~~~U(t,x)=\int^{+\infty}_{0}g(x')K(t,x,x')dx'+\int^{+\infty}_{0}f(x')\frac{\partial}{\partial t}K(t,x,x')dx'\\
+\lim_{\mu\rightarrow\frac{1}{2}}\frac{\Gamma(1-\mu)}{2\sqrt{\pi}\Gamma(\frac{3}{2}-\mu)}t^{2\mu}x^{-\alpha-\frac{1}{2}}\\ \times\int^{x+t}_{x-t}\frac{f(x')x'^{\alpha+\frac{1}{2}}}{x-x'}~_{2}F_{1}(\frac{1}{2}-\alpha,\frac{1}{2}+\alpha,\frac{3}{2}-\mu,X)\frac{d}{dx'}[t^{2}-(x'-x)^{2}]^{\frac{1}{2}-\mu}dx'$,\\ 
\\
une int\'egration par parties montre que la valeur de la derni\`ere limite est\\
\\
$\frac{1}{2}x^{-\alpha-\frac{1}{2}}[f(x-t)(x-t)^{\frac{1}{2}+\alpha}+f(x+t)(x+t)^{\frac{1}{2}+\alpha}]~~~~$, d'o\`u\\
\\
$U(t,x)=\int^{+\infty}_{0}g(x')K(t,x,x')dx'+\int^{+\infty}_{0}f(x')\frac{\partial}{\partial t}K(t,x,x')dx'\\
+\frac{1}{2}x^{-\alpha-\frac{1}{2}}[f(x-t)(x-t)^{\frac{1}{2}+\alpha}+f(x+t)(x+t)^{\frac{1}{2}+\alpha}]$.\\
\\
{\bf - Pour x $<$ t,} le changement des variables $x'=\sqrt{t^{2}-(1-z^{2})x^{2}}+zx,$ donne\\
\\
$U(t,x)=\int^{+\infty}_{0}g(x')K(t,x,x')dx'
+\int^{t-x}_{0}f(x')\frac{\partial}{\partial t}K(t,x,x')dx'\\ -\frac{1}{2}tx^{-\alpha-\frac{1}{2}}\lim_{\mu\rightarrow\frac{1}{2}}\int^{1}_{-1}\frac{f(x')x'^{\frac{1}{2}+\alpha}}{\sqrt{t^{2}-(1-z^{2})x^{2}}}\frac{d}{dz}[(1-z^{2})^{\frac{1}{4}-\frac{\mu}{2}}P^{\mu-\frac{1}{2}}_{\alpha-\frac{1}{2}}(z)]dz$,\\
\\
en utilisant la relation $[7]~P167$
$$P^{\mu-\frac{1}{2}}_{\alpha-\frac{1}{2}}(z)=\frac{1}{\cos(\mu-\frac{1}{2})\pi}[\frac{\Gamma(\alpha+\mu)}{\Gamma(\alpha+1-\mu)}P^{\frac{1}{2}-\mu}_{\alpha-\frac{1}{2}}(z)+\frac{2}{\Gamma(\mu-\frac{1}{2})\Gamma(\frac{3}{2}-\mu)}Q^{\mu-\frac{1}{2}}_{\alpha-\frac{1}{2}}(z)],$$

on voit que la valeur de la derni\`ere limite est\newpage
$\int^{t+x}_{t-x}f(x')\frac{\partial}{\partial t}K(t,x,x')dx'\\
-\frac{1}{2}tx^{-\alpha-\frac{1}{2}}lim_{\mu\rightarrow\frac{1}{2}}\int^{1}_{-1}\frac{f(x')x'^{\frac{1}{2}+\alpha}}{\sqrt{t^{2}-(1-z^{2})x^{2}}}\frac{d}{dz}[(1-z^{2})^{\frac{1}{4}-\frac{\mu}{2}}(\frac{2}{\Gamma(\mu-\frac{1}{2})\Gamma(\frac{3}{2}-\mu)}Q^{\mu-\frac{1}{2}}_{\alpha-\frac{1}{2}}(z))]dz$,\\
\\soit alors\\
$U(t,x)=\int^{+\infty}_{0}g(x')K(t,x,x')dx'+\int^{+\infty}_{0}f(x')\frac{\partial}{\partial t}K(t,x,x')dx'\\
-\frac{1}{2}tx^{-\alpha-\frac{1}{2}}lim_{\mu\rightarrow\frac{1}{2}}
[\frac{f(x')x'^{\frac{1}{2}+\alpha}}{\sqrt{t^{2}-(1-z^{2})x^{2}}}(1-z^{2})^{\frac{1}{4}-\frac{\mu}{2}}\frac{2}{\Gamma(\mu-\frac{1}{2})\Gamma(\frac{3}{2}-\mu)}Q^{\mu-\frac{1}{2}}_{\alpha-\frac{1}{2}}(z)]^{1}_{-1}\\
+\frac{1}{2}tx^{-\alpha-\frac{1}{2}}lim_{\mu\rightarrow\frac{1}{2}}\int^{1}_{-1}\frac{d}{dz}[\frac{f(x')x'^{\frac{1}{2}+\alpha}}{\sqrt{t^{2}-(1-z^{2})x^{2}}}](1-z^{2})^{\frac{1}{4}-\frac{\mu}{2}}\frac{2}{\Gamma(\mu-\frac{1}{2})\Gamma(\frac{3}{2}-\mu)}Q^{\mu-\frac{1}{2}}_{\alpha-\frac{1}{2}}(z)dz $,\\
\\
d'apr\`es le comportement asymptotique de la fonction $Q^{\mu}_{\nu}~[7]~P196-197$ \\
\\
$Q^{\mu}_{\nu}(z)\approx2^{-1-\frac{1}{2}\mu}\Gamma(-\mu)\frac{\Gamma(\nu+\mu+1)}{\Gamma(\nu-\mu+1)}(1-z)^{\frac{\mu}{2}}~pour~z\approx1,~\mu<0,\\
Q^{\mu}_{\nu}(z)\approx2^{-1-\frac{1}{2}\mu}\Gamma(-\mu)\cos[\pi(\nu+\mu)]\frac{\Gamma(\nu+\mu+1)}{\Gamma(\nu-\mu+1)}(1+z)^{\frac{\mu}{2}}~pour~z\approx-1,~\mu<0$,\\
\\ 
on a $~~~~~~[\frac{f(x')x'^{\frac{1}{2}+\alpha}}{\sqrt{t^{2}-(1-z^{2})x^{2}}}(1-z^{2})^{\frac{1}{4}-\frac{\mu}{2}}\frac{2}{\Gamma(\mu-\frac{1}{2})\Gamma(\frac{3}{2}-\mu)}Q^{\mu-\frac{1}{2}}_{\alpha-\frac{1}{2}}(z)]^{1}_{-1}=\\
\frac{1}{2}x^{-\alpha-\frac{1}{2}}[-\sin\pi\alpha.f(t-x)(t-x)^{\frac{1}{2}+\alpha}+f(t+x)(t+x)^{\frac{1}{2}+\alpha}]$;\\
\\
et d'apr\`es le th\'eor\`eme de convergence domin\'ee de Lebesgue on a\\ 
\\
$\lim_{\mu\rightarrow\frac{1}{2}}\int^{1}_{-1}\frac{d}{dz}[\frac{f(x')x'^{\frac{1}{2}+\alpha}}{\sqrt{t^{2}-(1-z^{2})x^{2}}}](1-z^{2})^{\frac{1}{4}-\frac{\mu}{2}}\frac{2}{\Gamma(\mu-\frac{1}{2})\Gamma(\frac{3}{2}-\mu)}Q^{\mu-\frac{1}{2}}_{\alpha-\frac{1}{2}}(z)dz=0$\\
\`a savoir que $\frac{2}{\Gamma(\mu-\frac{1}{2})\Gamma(\frac{3}{2}-\mu)}Q^{\mu-\frac{1}{2}}_{\alpha-\frac{1}{2}}(z)=P^{\mu-\frac{1}{2}}_{\alpha-\frac{1}{2}}(z)-\frac{1}{\cos(\mu-\frac{1}{2})\pi}\frac{\Gamma(\alpha+\mu)}{\Gamma(\alpha+1-\mu)}P^{\frac{1}{2}-\mu}_{\alpha-\frac{1}{2}}(z)$ d'o\`u \\
$~~~~U(t,x)=\int^{+\infty}_{0}g(x')K(t,x,x')dx'+\int^{+\infty}_{0}f(x')\frac{\partial}{\partial t}K(t,x,x')dx'\\
+\frac{1}{2}x^{-\alpha-\frac{1}{2}}[-\sin\pi\alpha.f(t-x)(t-x)^{\frac{1}{2}+\alpha}+f(t+x)(t+x)^{\frac{1}{2}+\alpha}]$.\\
\\
{\bf Corollaire 3} ( Th\'eor\`eme 2.1.1 [2] ).\\
Pour $\mu=\frac{1}{2},~\nu=\frac{k+1}{2}$ dans le th\'eor\`eme 3 bis, on retrouve la solution exacte de l'\'equation homog\`ene d'Euler-Poison-Darboux
$$U(t,x)=\sum^{\infty}_{l=0}a_{l}U_{l}+\sum^{\infty}_{l=0}b_{l}V_{l}$$
o\`u $U_{l}=x^{l}~_{2}F_{1}(\frac{-l}{2},\frac{k+1-l}{2},\frac{1}{2},\frac{t^{2}}{x^{2}})$ et $V_{l}=tx^{l}~_{2}F_{1}(\frac{-l}{2},\frac{k+1-l}{2},\frac{3}{2},\frac{t^{2}}{x^{2}})$.\\
{\bf Exemples}.\\ 
{\bf 1.} Le probl\`eme $\left\{
\begin{array}{rr}
(\frac{\partial^{2}}{\partial x^{2}}-\frac{3}{x}\frac{\partial}{\partial x})U(t,x)=\frac{\partial^{2}}{\partial t^{2}}U(t,x)\\
U(0,x)=0,~~U_{t}(0,x)=x\\
\end{array}\right.$\\
admet la solution unique $U(t,x)=t\sqrt{x^{2}-t^{2}}$.\\
\\
{\bf 2.} Le probl\`eme $\left\{
\begin{array}{rr}
(\frac{\partial^{2}}{\partial x^{2}}+\frac{1}{x}\frac{\partial}{\partial x})U(t,x)=\frac{\partial^{2}}{\partial t^{2}}U(t,x)\\
U(0,x)=x,~~U_{t}(0,x)=0\\
\end{array}\right.$\\
admet la solution unique $U(t,x)=\sqrt{x^{2}-t^{2}}+t\arcsin\frac{t}{x}$.\\
En perspective, on \'etudiera les \'equations d'Euler-Poisson-Darboux \`a conditions modifi\'ees dans les espaces hyperboliques et elliptiques.}\\
\\
\\
\\
\scriptsize{{\bf 6. R\'EF\'ERENCES BIBLIOGRAPHIQUES.}\\
${\bf[1]}$- J. Barros-Neto: Hypergeometric functions and the Tricomi operator, arXiv:math/0310480v1 [math.AP] 30 Oct 2003.\\
${\bf[2]}$- A.Bentrad: Exact solutions for a different version of the monhomogeneouse E-P-D equation.Complex variables and elliptic equations,vol.51.No.3 March 2006,243-253.\\
${\bf[3]}$- D.W.Bresters: On the equation of Euler-Poisson-Darboux.Siam J.Math.Anal.1973 no.1, 31-41.\\
${\bf[4]}$- L.Colzani: Radial solutions to the wave equation. Annali di matematica 181, 25-54 (2002).\\
${\bf[5]}$- I.S. Gradshteyn and I.M. Ryzhik: Table of Integrals, Series, and Products; sixth edition. Academic press 2000.\\
${\bf[6]}$- N.N.Lebedev: Special functions And their applications. Dover Publications,Inc New york 1972.\\ 
${\bf[7]}$-  W. Magnus, F. Oberhettinger, and R. P. Soni: Formulas and Theorems for the special Functions of Mathematical Physics, Springer-Verlag, New York, 1966.\\
${\bf[8]}$- Raimundas Vidunas: Specialization of Appell's functions to univariate hypergeometric functions. J. Math. Anal. Appl. 355 (2009) 145-163.\\
${\bf[9]}$- A. Weinstein: On the wave equation and the equation of Euler-Poisson, proc. Symposia Appl. Math, vol. 5, McGraw-Hill, New York, 1954, pp. 137-147.}\\
\begin{flushleft}
\tiny{\bf Universit\'e Gaston Berger de Saint-Louis B.P: 234. S\'en\'egal.\\
E-mail adresse: cheikh976@yahoo.fr.\\
Universit\'e de Nouakchott Facult\'e des sciences et techniques B.P: 5026. Mauritanie.\\
E-mail adresse: mohamedvall.ouldmoustapha230@gmail.com.}
\end{flushleft}
\end{document}